# Enthalpy and high temperature relaxation kinetics of stable vapor-deposited glasses of toluene

Deepanjan Bhattacharya and Vlad Sadtchenko[*]
The George Washington University
Chemistry Department
Washington, DC
Corresponding author email: vlad@gwu.edu

**ABSTRACT:**

Stable non-crystalline toluene films of micrometer and nanometer thicknesses were grown by vapor deposition at distinct rates and probed by Fast Scanning Calorimetry. Fast scanning calorimetry is shown to be extremely sensitive to the structure of the vapor-deposited phase and was used to characterize simultaneously its kinetic stability and its thermodynamic properties. According to our analysis, transformation of vapor -deposited samples of toluene during heating with rates in excess $10^5$ K·s$^{-1}$ follows the zero-order kinetics. The transformation rate correlates strongly with the initial enthalpy of the sample, which increases with the deposition rate according to sub-linear law. Analysis of the transformation kinetics of vapor deposited toluene films of various thicknesses reveal a sudden increase in the transformation rate for films thinner than 250 nm. The change in kinetics correlates with the surface roughness scale of the substrate, which is interpreted as evidence for kinetic anisotropy of the samples. The implications of these findings for the formation mechanism and structure of vapor deposited stable glasses are discussed.

**I. INTRODUCTION**

The discovery of highly stable glasses prepared by physical vapor deposition stimulated intense experimental and theoretical efforts to elucidate structure and properties of these remarkable materials [1-21]. According to the estimates of the fictive temperatures of the novel phases, an attempt to achieve similar kinetic stability and low enthalpy by traditional glass annealing would require times almost on a geological scale. Current consensus on the mechanisms of formation of such stable glasses emphasizes the role of molecular dynamics at the free surface of a growing sample.

In the framework of the potential energy landscape theory of glass transition, formation of an ordinary glass occurs when the temperature of the melt is lowered below certain value (standard glass transition temperature, $T_g$) at which the liquid is no longer capable of overcoming the inter-basin barriers, and thus become "trapped" in a glassy configuration[13, 22]. Such a configuration does not necessarily represent a global minimum of potential energy of the system. However, further reduction in energy and the resulting increase in kinetic stability may require exceedingly long times. Unlike the ordinary preparation technique, vapor deposition of glasses is thought to take advantage of high rates of constituent diffusion along the surface of the growing sample. The high mobility of molecules at the vapor-glass interface [10] makes it possible to access the lowest states on the potential energy landscape on a typical laboratory time scale.

In early experiments, the successful preparation of the high stability, low enthalpy phase was typically observed when the deposition rates were low (sub-nanometer per second) and the temperatures were approximately 0.86 of the standard $T_g$ [1, 3]. From the view point of the surface mobility hypothesis, such interplay between the deposition rate and the deposition temperature is not surprising. For a certain temperature, the deposition rate has to be sufficiently low so that constituents would have sufficient time to diffuse along the sample surface and find a minimum energy configuration before being trapped under subsequent molecular layers.

Although providing a reasonable explanation for formation of high stability phase, the mechanism described immediately above has not been completely validated. Results of recent studies of vapor deposited glassy films of toluene and ethylbenzene seem to fall out from the simple picture of surface mediated relaxation. For example, studies of kinetic stability of vapor deposited films of different compound produced conflicting result on the impact of the deposition rate on properties of samples[2, 15]. Leone-Gutierrez et. al. [5] reported that the kinetic stability of vapor-deposited ethylbenzene decreases in the limit of low deposition rates, which is in contrast to finding of Ahrenberg et. al.[15], and Ramos et. al.[9], and which is difficult to explain in the framework of the current formation mechanism conjecture. Furthermore, several experimental observations are not properly explained by current empirical scenario of stable phase formation which postulate that high surface diffusion rates of the phase constituents allow much faster relaxation of the sample to a low enthalpy state which, in principle, can be achieved by aging the ordinary glass over extraordinary long times. Indeed, past wide angle X-ray scattering (WAXS) studies revealed structural anisotropy in stable vapor-deposited tris-naphthylbenzene films[12]. More recent WAXS and computational investigations of stable toluene samples by Ishii and Nakayama also imply that the structure of vapor-deposited phase of toluene may be distinct from the ordinary glass, and may consist of locally stable aggregates [21]. Finally, the formation of anisotropic (layered) stable phase was also observed in computational studies of the structure of a vapor-deposited glass of trehalose[7].

With the objective of clarifying the formation mechanism, molecular structure, and kinetic stability of vapor-deposited and ordinary glasses, we initiated a Fast Scanning Calorimetry (FSC) study of selected low temperature organic glasses. In our experiments, micrometer scale toluene films of maximum kinetic stability were prepared at temperature near 112 K. The transformation of such films into ordinary supercooled liquid was monitored during their heating with the rates in excess of $10^5$ K·s$^{-1}$. Under such extreme conditions, the transformation takes place at temperatures significantly higher than previously observed (up to 150 K). Furthermore, due to high heating rates, the transformation occurs over relatively broad temperature range, making it possible to gain insights into temperature dependence of the transformation rates at these high temperatures. In this article, we focus on the mechanisms of transformation of stable non-crystalline phase of toluene formed by vapor deposition at distinct rates.

## II. EXPERIMENTAL

### A. Fast Scanning Calorimetry.

The central idea of FSC technique consists in measuring heat capacity of a glassy or liquid sample during heating from an initial state with rates ranging from $10^3$ to $10^6$ K·s$^{-1}$, i.e., up to $10^8$ times higher than those employed in traditional Differential Scanning Calorimetry (DSC) experiments. The primary advantage of the high heating rate is the high temperature of relaxation of liquid or glassy sample during the scan. For example, a glassy film of toluene heated with a rate on the order of $10^5$ K·s$^{-1}$ relaxes to equilibrium at temperatures near 135 K., *i.e.*, at temperatures almost 20 K above the standard glass transition temperature (117 K). As illustrated in the upper panel of Fig. 1, the return to equilibrium during a rapid scan is manifested in the FSC thermograms by large and sharp endotherms, which onset temperature and magnitude are highly sensitive to sample preparative conditions. Because the kinetics of high temperature relaxation during fast scan must depend on the initial structural and thermodynamic state of the sample, the fast relaxation endotherms (FREs) can be used to probe the sample properties in a wide range of temperatures from those below and *significantly above* the standard glass transition.

The outline of the experimental apparatus for FSC studies is shown in the lower panel of Fig. 1[23]. The "sample holder" is essentially a 10 μm in diameter, 1.5 cm long tungsten filament attached to supports inside a vacuum chamber maintained at a pressure of approximately 5·10$^{-7}$ Torr. The temperature of the filament during film's preparation is controlled by adjusting the temperatures of the supports. Each of the supports is equipped with a 10 W heater and a miniature T-type thermocouple. Two feedback controllers are used to vary and maintain preset temperature values of each support, *i.e.*, the temperature of each support is adjusted independently. However, the actual temperatures of the supports are measured by a single temperature monitor, which has two thermocouple inputs and is capable of measuring *the difference* between support temperatures with overall accuracy of 0.2 K. The accuracy of this comparative measurement was verified by registering negligible difference between the temperatures of two supports when the entire apparatus was at room temperature. Such an arrangement makes it possible to minimize longitudinal temperature gradient in the tungsten filament by adjusting the controller's set point and monitoring the difference in the temperatures of the supports. It also makes it possible to gauge the impact of such gradients on the results of FSC experiments by intentionally setting the temperatures of the supports to distinct values. We discuss the temperature calibration procedure in greater detail in Sec. II B of this article.

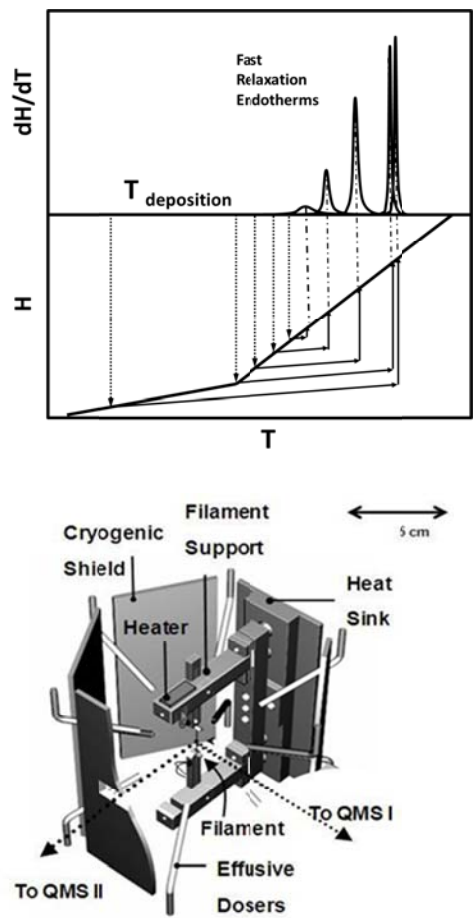

**Figure 1.** *Upper panel:* Fundamentals of Fast Scanning Calorimetry. Vapor deposition of constituents on a cold substrate may result in formation of non-crystalline samples characterized by distinct enthalpies. Rapid heating of such samples results in fast relaxation endotherms (FREs) at temperatures tens of degrees above the initial temperature. The position and shape of FREs are sensitive to the initial state of the sample and can be used to gain information on relaxation kinetics. Integration of the thermograms provides insights into initial thermodynamic state. *Lower panel:* Outline of the fast scanning calorimeter. See text for details.

As shown in the lower panel of Fig. 1, films of volatile material are deposited on the surface of the filament via 12 effusive dosers surrounding the filament. The dosers are essentially stainless steel tubes of 1/8" in diameter. The dosers are connected to a vapor source of our design positioned outside the main chamber. The deposition time can be adjusted from 0.05 s to several hours. The deposition rates can be adjusted from 0.5 to 50 nm·s$^{-1}$.

After film preparation, fast heating of the filament is initiated by applying a potential difference across the filament. The data acquisition (DAQ) system of the apparatus simultaneously measures the voltages drop across the filament and the current through the filament. These data are used later to calculate the filament resistance and the dissipated power. Two current measurements and two voltage measurements are collected every 4 microseconds. The temperature of the filament is inferred from its resistance using data from a calibration procedure described in Sec. II B, *i.e.*, the filament is simultaneously a heater and a temperature sensor. Heat capacity of the filament is calculated as the ratio of the power dissipated by the filament to the first time derivative of its temperature. The heat capacity of a film on the filament surface is determined by subtracting heat capacity of the bare filament from the heat capacity of the filament covered by the film.

In order to avoid temperature gradients due to radiative cooling or heating of the filament, the filament assembly is surrounded by six cryogenic shields maintained at approximately 105 K. In order to verify that radiative cooling or heating of the filament during deposition is negligible, we measured resistance of the filament while varying the temperature of both supports in range from 100 to 160 K. Because, in the case of significant radiative cooling, the greater difference between the temperature of the supports and the surrounding cryogenic shields must result in progressively greater difference between the temperature of the filament and that of the supports (due to progressively higher radiation flux from the filament to shields), a non-linear dependence of the filament's resistance on support temperature is expected. Nevertheless, no significant deviations from linearity were observed when the resistance of the filament was plotted against the average support temperatures. According to our estimates, the impact of radiative cooling on the temperature of the filament is significantly less than 1 K in the range of temperatures from 100 to 160 K.

### C. Temperature calibration

The following procedure was used to calibrate the FSC apparatus for temperature measurements during fast heating stage of the experiments and during deposition. The temperature of the support was set at a value between 100 and 180 K and the resistance of the filament was measured with an ohmmeter. The temperature of the filament was assumed to be equal to the temperature of the supports. The temperature values were plotted against the resistance and fitted with a linear function. The slope and intercept values were later used to calculate filament temperature from filament resistance data measured during rapid heating.

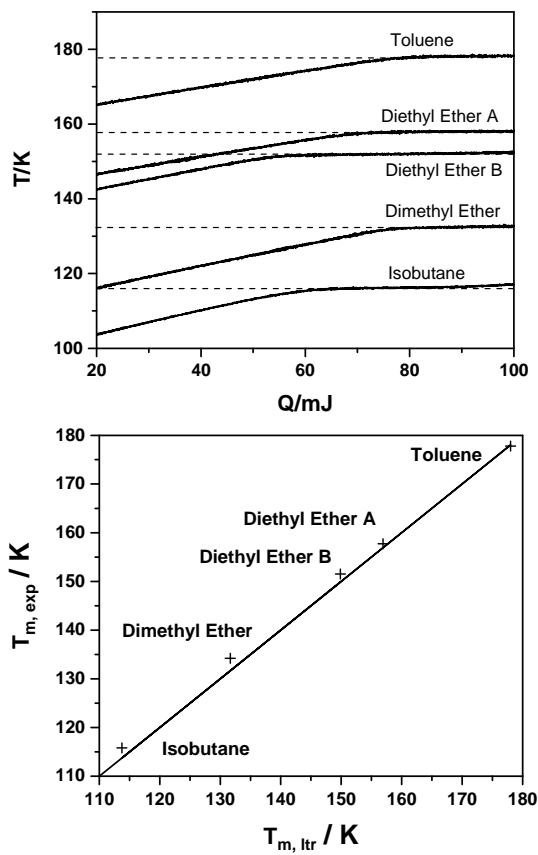

**Figure 2.** *Upper panel:* Heating curves of five different crystalline films prepared under conditions where vapor deposition resulted in crystallization upon deposition during film growth. The film thicknesses were on the order of a micrometer. The initial heating rate is near 2000 K/s. *Lower panel:* The melting temperatures inferred from the heating curves as a function of the literature values. Deviation from the solid line represent systematic offset of the calibration. See text for details.

The overall accuracy of temperature measurements was verified by two types of tests. In the first series of tests, micrometer scale films of crystalline toluene, two allotropes of diethyl ether, crystalline dimethyl ether, and crystalline isobutane were prepared by direct vapor deposition at temperatures several Kelvins below the respective melting points of the compounds. After preparation, the films were subjected to heating with the rate on the order of 2000 K·s$^{-1}$, and the heating curves, *i.e.*, plots of film temperature as a function of heat generated by the filament were obtained.

The heating curves for five crystalline films with melting temperatures ranging from 113.5 to 178.8 K are summarized in the upper panel of Fig. 2. As shown in the figure, the onset of melting is manifested as a sudden change in the slope of the curve at a particular temperature.

The lack of significant temperature variations with the heat generated in the thermodynamic system is the classic sign of a first order phase transition, *i.e.*, melting of the crystalline films. The temperatures of the plateaus observed in heating curves were used for verification of temperature measurements in our experiments.

Lower panel of Fig. 2 summarizes the results of analysis of the heating curves. It compares the melting temperatures derived from the plots in the upper panel of Fig. 2 with those found in literature[24-27]. The comparison of previously established melting temperatures and those determined in our measurements revealed a *systematic* offset in our temperature measurements of approximately +2.5 K in the range from 110 to 150 K. The calibration was adjusted to compensate for this positive offset.

The temperature interval from 110 to 150 K is critical for interpretation of FSC results on relaxation in toluene films described in this article. Therefore, we invested considerable effort to elucidate the origins of this systematic error in temperature calibration. According to our tests, the offset arises from slight increase (0.4 %) in the baseline resistance of the filament upon deposition of films of molecular solids. Because the temperature calibration procedure was based on measurements of *bare* filament resistance at various temperatures, the results of the linear fit of resistance values produced systematic positive offset in the temperature of the filament covered by molecular films. Although the physical origins of these phenomena are unclear to us, we have determined that the magnitude of the positive offset is independent of either the thickness of the film or its chemical composition. We speculate that the slight increase in the resistance of the filament is due to enhanced electron scattering at the interface between filament and molecular films, which represents important contribution to the overall resistivity in thin wires[28].

The melting curves in Fig. 2, were measured by heating a crystalline films to its melting temperature with relatively high rates (approximately 2000 K·s$^{-1}$), which poses concerns about possible crystal overheating and temperature lags during melting. Indeed, careful analysis of the melting curves shows that the first time and energy derivatives of the temperature in these experiments are finite during melting, and that the temperature may increase by up to 1.5 K as the melting progresses. Note that the transition temperatures derived in such "dynamic melting" experiments are those characteristic of the *onset* of the melting processes.

In order to explore potential differences between phase transition temperatures observed in the dynamic melting experiments and those determined using DSC and other conventional techniques, we conducted systematic investigations of crystallization of vapor deposited films under isothermal conditions. In these experiments, the temperature of the filament assembly was brought to a value near material's melting point, and a micrometer scale film was directly deposited on the surface of the filament. Immediately after deposition the heating of the film was initiated to determine the phase composition of the films formed under isothermal conditions. The results of these experiments are summarized in Fig. 3.

Upper panel of Figure 3 shows heating curves of micrometer scale isobutane films deposited at distinct temperatures. As illustrated in Fig. 3, the heating curves of films, which are devoid of crystallites, show no change in slope over the entire heating

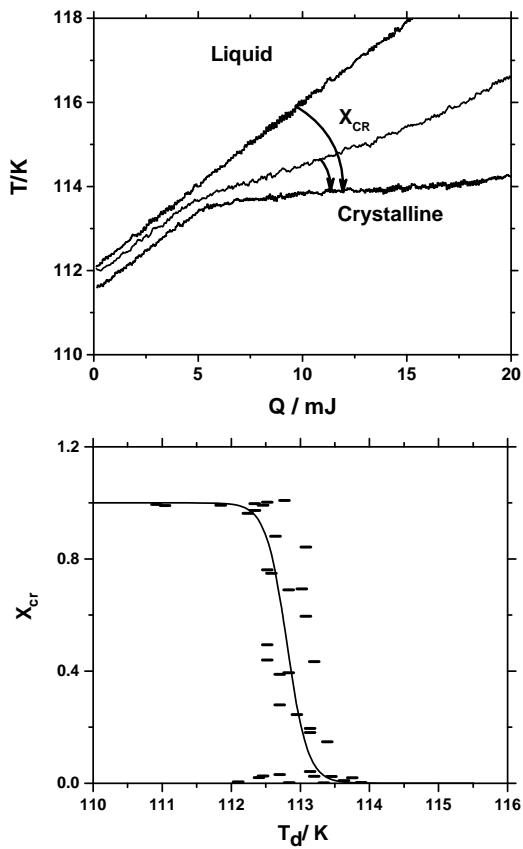

**Figure 3.** *Upper panel:* Heating curves of micrometer scale isobutane vapor deposited at temperatures above and below its melting point. The deposition temperature (as read by DAQ system) is inferred by extrapolating the heating curve to zero heat. The changing slope of the heating curves signifies the extent of sample crystallinity (see text for details).

*Lower panel:* The crystalline fraction of isobutane films vapor deposited at distinct temperatures, $T_d$. The rapid decrease in crystalline fraction signifies approach to melting temperature of the compound. The solid line is an eye guide, not a fit.

range (see the curve labelled "Liquid" in Fig. 3). As the deposition temperature decreases, the resulting heating curves gradually assume the shape of fully crystallized samples (see the curve labeled "Crystalline" in Fig. 3). These curves are characterized by a plateau near the melting temperature of isobutane observed during rapid heating. The intermediate case is consistent with formation of a partially crystalline film. The fraction of the film that crystallizes during deposition ($X_{CR}$) can be estimated from the differences in slopes of the heating curves of either liquid or crystalline film and those of the partially crystalline film. The central assumption of this determination was that the heating curve of a partially crystallized film is a linear combination of the heating curves of the fully liquid and fully crystalline samples.

The lower panel of Fig. 3 summarizes results of analysis of approximately 40 heating curves of isobutane films deposited at distinct temperatures under near isothermal conditions. It shows the extent of crystallization, *i.e.*, the estimated crystalline fraction, $X_{CRr}$, as a function of the deposition temperature.

As shown in the figure, formation of fully crystalline, partially crystalline, and fully liquid films is possible in the temperature range from 112 K to 113.5 K. Random fluctuations of $X_{CR}$ values in this temperature range are fully consistent with stochastic nature of crystal nucleation process under conditions of moderate supercooling and frequently observed formation non-equilibrium (supercooled) liquid[29]. Nevertheless, the probability of formation of even a partially crystalline film becomes negligible at temperatures above 113.5 K which is within tenth of a degree from the triple point of isobutane[27]. Because formation of superheated crystalline phase is extremely unlikely under near isothermal condition, we conclude that the temperature calibration procedure used in our experiments is correct and that the overall systemic accuracy of deposition temperature measurements is within fraction of a degree in the temperature range near 113.5 K

Note also that the width of transition from fully crystalline to fully liquid state of the sample observed in our near-equilibrium experiments (see the lower panel of Fig. 3) represents the upper estimate for temperature gradient along the filament during deposition. Based on the analysis of the data summarized in Fig. 3, we conclude that the longitudinal temperature gradients during film deposition are less than 1 K.

**C. Choice of heating rate and film thickness: temperature lags and gradients.**

High heating rates utilized in our FSC studies rise concerns about a potential impact of transverse temperature lags and gradients on the results of FSC experiments with vapor deposited and ordinary supercooled films of toluene. We emphasize that the choice of the maximum thickness and the maximum heating rates employed in our experiments is not arbitrary and is based on maximizing the signal to noise ratio and the heating rate while keeping potential thermal lags below a value which could impact the heat capacity measurements. The criteria, used to ensure acceptable thermal lag across the film, were the thermal equilibration time $\tau_{vib}$, and the characteristic experimental time $t_{exp}$, which is 1K divided by the maximum heating rate (10 microseconds). The clear condition for avoiding significant thermal lags during heating is to ensure that $t_{exp}$ exceeds the thermal relaxation time, $\tau_{vib}$.

An estimate of $\tau_{vib}$ for a toluene film on a surface of a cylindrical filament can be obtained using Einstein-Smoluchowski equation, $\tau_{vib} \approx L^2/4D_t$, where $L$ is the thickness of the film, and $D_t$ is the thermal diffusivity. The thermal diffusivity of liquid toluene at 180 K is approximately $10^{-7}$ m$^2 \cdot$s$^{-1}$ [30], which gives the limiting value for the sample thickness of approximately 2 μm for heating rates typically employed in our FSC experiments ($10^5$ K·s$^{-1}$). The impact of possible thermal lags and gradient was experimentally verified by changing the films thickness and the heating rate. According to these tests the impact of thermal lags on FSC thermograms is negligible as long as the film thickness is below 2 μm.

## III. RESULTS

### A. Selected FSC thermograms of toluene films vapor deposited at distinct temperatures.

Fig. 4 shows selected results of FSC studies of toluene films vapor- deposited on the surface of the filament at temperatures between 95 and 155 K. The representative FSC thermograms were obtained for the films vapor-deposited at 112 K, 117 K, 125 K, and 148 K. The deposition rate was approximately 15 nm·s$^{-1}$. Each film was deposited for 120 s, which resulted in average film thickness of 1.8 μm.

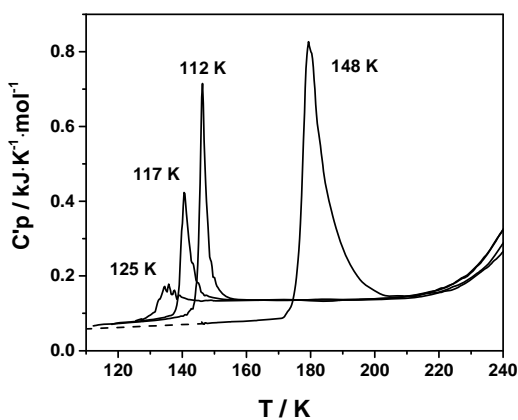

**Figure 4.** Representative FSC thermograms of toluene films vapor deposited at 148, 125, 117, and 112 K. Vapor deposition at temperatures above 148 K results in crystalline films which thermograms are characterized by a melting endotherm near 178 K. Lowering of temperature to 125 K leads to FSC thermograms characterized by a weak, low temperature endotherm near 130 K due to fast relaxation of ordinary supercooled liquid, which is fallen out of equilibrium due to rapid heating. As the temperature lowered further, the FREs shift to lower temperature, which signifies formation of films with lower enthalpy and higher kinetic stability. The films of maximum stability are prepared at deposition temperatures near 112 K. Further decrease in deposition temperature results in shift of FREs to higher temperatures.

As shown in the figure, the FSC thermogram of the toluene film vapor deposited at 112 K is characterized by a sharp endotherm with the onset near 145 K. The fast relaxation endotherm near 145 K signifies return to equilibrium of the rapidly heated toluene sample. Note that toluene samples never crystallize during rapid scans. Indeed, no exotherms are present in FSC thermograms at temperatures above the onset of FREs.

Increase in deposition temperature leads to a rapid decrease in the onset temperature of FRE, and a rapid decrease in the FRE's magnitude (see Fig. 4). When the deposition approaches a value near 130 K, the endotherms "disappear" from the FSC thermograms of toluene (not shown in Fig. 4). The lack of observable endotherms in the thermograms of toluene deposited at temperatures above 130 K is consistent with near-equilibrium enthalpy relaxation times of toluene approaching the characteristic observation time of the FSC experiments, $t_{exp}$ ($10^{-5}$ s). Indeed, the relaxation time of toluene is $10^{-5}$ s near 130 K [31, 32]. Therefore, heating rates in excess of those employed in our experiments are necessary for the sample to deviate significantly from equilibrium during the rapid scan.

At 148 K, vapor deposition leads to complete crystallization of toluene samples, which is manifested in the FSC thermogram as the melting endotherms near 178 K. The onset temperature of the melting endotherm observed in the FSC experiment is consistent with literature value for the melting of bulk-like samples[33]. Note that unlike the fast relaxation endotherms, the onset temperatures and magnitude of melting endotherms are independent of the deposition conditions. The strong dependence of FREs on the deposition condition, and the lack of such in the case of melting endotherms is a textbook example of fundamental distinctions between glass softening (or liquid relaxation) from typical first order phase transitions[13, 34, 35].

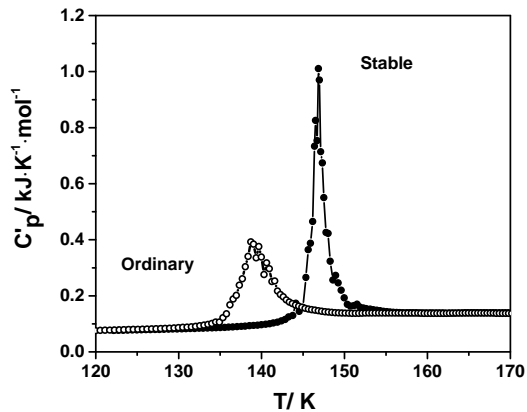

**Figure 5.** Comparison of fast relaxation endotherms of the most stable vapor-deposited toluene film with that of an ordinary glass prepared by slow cooling of melt from 127 to 112 K. The onset temperature of FRE of vapor deposited sample is 10 K higher than that of the ordinary glass which signifies a greater kinetic stability of the vapor-deposited film.

Analysis of the FSC thermograms of toluene films deposited at various temperatures showed that, the most stable films of vapor deposited toluene were formed at temperatures between 112 and 111 K. The decrease in temperature below 112 K resulted in gradual decrease in stability of films manifested by decrease in $T_{FRE}$ (not shown in the figure).

Figure 5 compares the FSC thermograms of a toluene films vapor deposited at temperature near 112 K (open circles) with that of a film which was vapor deposited at 127 K, and then slowly cooled to 112 K (solid circles). From now on, we will refer to samples prepared by slow cooling as "ordinary glass". The cooling rate during preparation of the ordinary toluene glass was 3.6 K·min$^{-1}$, *i.e.*, it was similar to the cooling rate typical of traditional DSC experiments. As shown in the Fig. 5, the onset temperature of the fast relaxation endotherm in the case of the vapor deposited sample is almost 10 K higher than that in the case of the case of the ordinary glass.

Vapor-deposited films of heightened stability were already tested by pulsed nanocalorimetry. For example, Sepulveda *et. al.* conducted studies of vapor deposited film softening kinetics at temperatures significantly above standard glass transition by subjecting the sample to rapid heating with the rate on the order of $3.5 \cdot 10^4$ K·s$^{-1}$ [36]. Those experiments, however, employed *significantly lower* vapor deposition rates and deposition temperatures compared to our experiments. In order to show that the vapor-deposited films of toluene prepared in our experiments have kinetic stability comparable to that of the samples prepared at lower temperatures and deposition rates, we measured FSC thermograms of vapor deposited toluene films at a heating rate of approximately $3.5 \cdot 10^4$ K·s$^{-1}$. Judging by the onset temperature of the FREs measured at lower heating rate, the kinetic stability of the samples prepared in our laboratory is similar to that of samples prepared by Sepulveda *et. al.*.

In addition to studies of transformation kinetics of stable toluene films under isothermal conditions, Sepulveda *et. al.*, provided estimates of the transformation times obtained from pulsed calorimetry thermograms measured at different rates. The transformation times, were estimated using Deborah criterion [37], *i.e.*, under assumption that the relaxation time at $T_{FRE}$ is numerically equal to the inverse of the heating rate. Although useful, such an estimate may not be sufficiently stringent [38]. Indeed, the Deborah, and more strict Lillie numbers based criteria were tested under conditions of slow cooling or heating, i.e., under near-equilibrium conditions[38]. Furthermore, it is unclear what such an estimate of far out of equilibrium relaxation time may represent, *i.e.*, how it is related to liquid or glassy relaxation times typically measured under near equilibrium conditions[37]. Finally note that the relationship between characteristic glass transition temperature and sample relaxation time were established in the case of ordinary glass and may not be valid in the case of vapor-deposited glasses of high stability due to their potentially distinct relaxation kinetics and mechanisms.

Taking into account uncertainties in the relationships between fast relaxation kinetics of glassy samples and their thermodynamic and structural properties, we attempted a more conventional test of the stability of our vapor-deposited toluene films. The results of this experiment are illustrated in Fig. 6. It shows a thermogram of a stable vapor deposited film, which was slowly warmed from 112 K to approximately 130 K. The heating rate during the slow warming was approximately 0.5 K·s$^{-1}$.

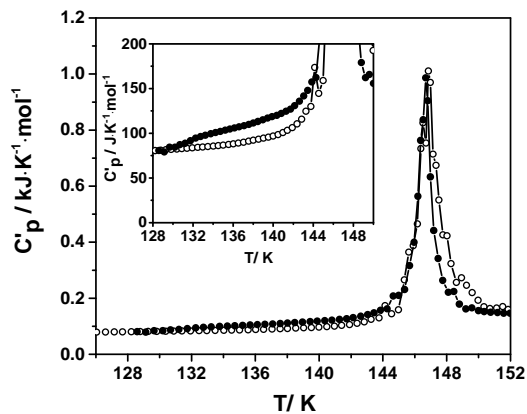

**Figure 6.** Comparison of the FSC thermograms of a toluene film vapor-deposited at 112 K with that of a film which was vapor deposited at 112 K and then heated with a rate of 0.5 K/s to 128 K. The decrease in the width of the endotherm and the increase in the heat capacity at temperatures below the onset of the FREs are consistent with partial transformation of the stable film into ordinary supercooled liquid during slow heating. See text for discussion.

As shown in Fig. 6, warming of the vapor deposited sample does not change the onset temperature of the endotherm. However, annealing results in a decrease in the width of the FRE, and a concurrent increase in heat capacity of the sample prior to fast relaxation (see inset in Fig. 6). Such a behavior is consistent with a *partial* transformation of the stable glass into ordinary liquid. When the fast scan initiated at 128 K, the resulting thermogram must be a linear combination of heat capacity of the ordinary liquid and that of the still stable untransformed fraction. Because near equilibrium relaxation times in ordinary liquid toluene are approaching the characteristic time scale of FSC experiment ($10^{-5}$ s) [31, 32] near 128 K, the contribution of ordinary liquid toluene fraction is simply an increase in the heat capacity prior to the onset of fast relaxation of untransformed stable fraction.

*In summary*, fast scanning and slow annealing experiments show that the micrometer scale films vapor-deposited at 112 K with the rate of 15 nm·s$^{-1}$ are significantly more stable than ordinary glassy or liquid samples. For example, the characteristic life time of the stable vapor deposited toluene is *at least* a second near 128 K. This value is drastically longer than a typical near equilibrium relaxation times in liquid toluene at the same temperature, which is on the order of $10^{-5}$ seconds. These findings are consistent with the conclusion of past AC calorimetry studies of thin toluene films[15], and with past FSC experiments[36]. While the vapor deposited toluene film discussed in this section may not represent the most stable non-crystalline toluene phase ever prepared (due to relatively high deposition rates), the chosen conditions are suitable to examine certain aspects the formation mechanisms and kinetics of softening of such samples.

## B. Transformation kinetics of stable toluene into ordinary liquid under conditions of rapid heating.

Ahrenberg *et. al.* recently showed that transformation of a stable vapor-deposited toluene films into ordinary liquid follows zero-order kinetics[15]. The softening begins at the surface of the films, and extends into the bulk of the film via a transformation front. Although the front velocity was found to be constant during a transformation under isothermal conditions, the front velocity is a strong function of temperature. Indeed, our analysis of the transformation kinetics data published by these researchers, showed that the zero-order transformation rate constant can be approximated by Arrhenius dependence on temperature with the effective activation energy of $180 \pm 20$ kJ·mol$^{-1}$ in the temperature range from 118 to 123 K. In order to gain insights into transformation kinetics of our stable toluene films, we developed a simple model of apparent heat capacity changes during fast relaxation in FSC experiments.

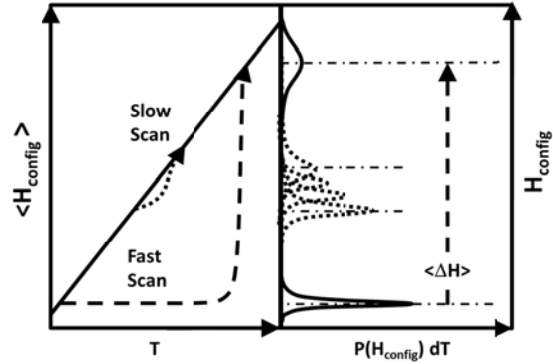

**Figure 7.** Illustration of fundamental difference between FSC and traditional DSC. Due to high heating rates fast relaxation in FSC experiments takes place at temperatures which are significantly higher than the starting temperature of the scan. As argued in the text, this leads to a significant separation between the enthalpy distribution of the initial (stable vapor-deposited glass) and final (ordinary liquid ), which greatly simplifies kinetics of the relaxation in the case of rapid scans.

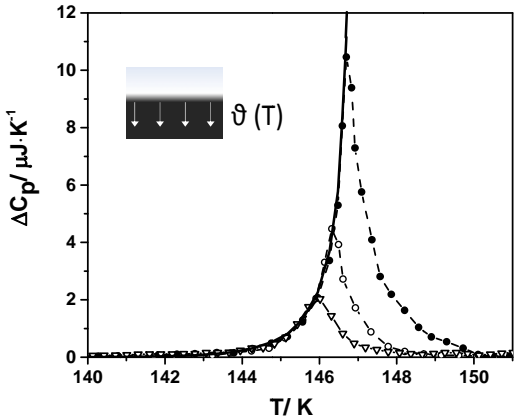

**Figure 8.** Excess heat capacity values during transformation of toluene films of different thicknesses deposited at distinct temperatures. The excess heat capacity peaks of 0.5 (open triangles), 1(open circles), and 2 (solid circles) μm thick films vapor-deposited at 112 K show the leading edge overlap representative of zero-order relaxation kinetics. The solid line shows the fit of excess heat capacity with a model, which assumes that the transformation to ordinary liquid begins at the film surface and progresses into the film's bulk via a transformation front which velocity, is constant at a particular temperature during the transformation. Dashed line represents the excess heat capacity of films deposited under conditions where a breakdown of zero-order kinetics is expected. See text for discussion.

The general assumption of the model is that, at a particular temperature, the excess heat capacity of the overshot peak, $\Delta C_p (T)$ or *(dH/dT)*, in the FSC thermograms of stable toluene samples can be related to the rate of transformation of the stable phase, *(dV/dt)*, via a simple relationship

$$\left(\frac{dH}{dT}\right) = \left(\frac{dV}{dt}\right) \cdot \left(\frac{dt}{dT}\right) \cdot \Delta H_{tr} \cdot n \ , \qquad (1)$$

where, *V* is the volume of the sample which has undergone the transformation, *(dt/dT)* is the inverse of the heating rate, *n* is the number density of the ordinary supercooled toluene, and $\Delta H_{tr}$ is the average enthalpy change per constituent during transformation, which is constant during the transformation.

The later assumption may seem farfetched under conditions of conventional DSC up-scan where a system does not deviate too far from equilibrium. However, it may be quite reasonable under conditions of fast heating of highly stable glasses where a system can be significantly far from equilibrium prior to relaxation. As illustrated in Fig. 7, in a slow scanning calorimetry experiment with an unaged glass, the distribution of enthalpies evolves continuously during the glass softening transition resulting in complex kinetics[37, 39]. The gradual molecular rearrangement is taking place throughout the bulk of the sample and the differences in distribution on initial (glassy) and final (liquid) states are blurred. The meaning of neither the volume of transformation nor the enthalpy recovered is clear.

In the limit of high heating rates, however, the system quickly falls far out of equilibrium which results in great separation between enthalpy distributions of rapidly heated stable glass sample and that of an equilibrium liquid at the temperature prior to the transition. Due to a clear distinction between the distribution of enthalpies of initial and final states in the limit of high heating rates, the transformation kinetics are likely to simplify. Indeed, the volume of transformed stable phase is the volume of nearly equilibrated supercooled liquid toluene, and $\Delta H_{tr}$ is the average enthalpy jump up on relaxation. Because the average enthalpy $\Delta H_{tr}$ is large, the variations in the enthalpy of the final (ordinary liquid) and initial (unrelaxed) states over the temperature range of the transition can be neglected. In short, Eq. 1 may capture essential features of the transformation kinetics, at least in the early stages of the transition.

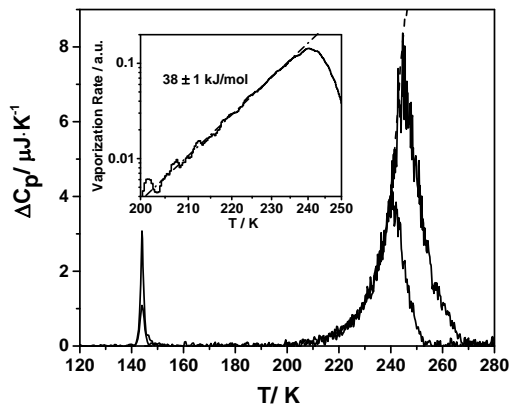

**Figure 9.** Excess heat capacity of 0.5 and 1 μm thick films of toluene in the temperature range from 120 to 280 K. The large endotherm with onset temperatures near 200 K represent the vaporization of ordinary liquid films of toluene, which is a classic example of a process which follows zero-order kinetics. Note the characteristic overlap of the leading edges of the vaporization endotherms. The inset shows the vaporization rate values at various temperatures during vaporization plotted in Arrhenius coordinates. The activation energy for vaporization derived from the thermograms is in agreement with the literature values of the enthalpy of vaporization of liquid toluene.

Figure 8 shows the result of the model fit (solid line) to the fast relaxation endotherms of most stable toluene films with thickness of approximately 1.8, 0.9, and 0.5 μm (closed and open circles, and open triangles circles respectively). The model assumes that the transformation rate is zero-order, *e.g.*, that the transformation of stable toluene film into ordinary liquid starts at its surface and progresses into the bulk with a constant velocity at a particular temperature. In this case, the derivative of volume in Eq. 1 is the transformation front velocity multiplied by the surface area of the film. Note the overlap of the leading edge of the endotherms which is typical for any temperature scanning experiment on a zero-order process. Indeed, if the y-axis title is hidden, the endotherms in Fig. 8 may be mistaken for Temperature Programed Desorption spectra of multilayer films of distinct thicknesses[40].

In order to demonstrate that the model represented by Eq. 1, correctly describes the shapes and positions of endotherms characteristic of a zero-order processes observed in FSC experiments, we used it to simulate the vaporization of toluene films at temperatures above 200 K.

Figure 9 shows the excess heat capacity of toluene films in the temperature range from 120 to 280 K. After relaxation into ordinary supercooled liquid near 145 K, the films begin to desorb rapidly from the filament (substrate) at temperatures above 200 K. Because the diameter of the filament is less than the mean free path of toluene molecules in saturated toluene vapor in the temperature range from 200 to 245 K, the vaporization into vacuum must occur without collisions, i.e., the vaporization process is not limited by diffusion in the vapor phase[41].

Under isothermal conditions, desorption of a multilayer film of a volatile material into vacuum must follow zero order kinetics with Arrhenius rate constant [40-42]. The activation energy for vaporization is the enthalpy of vaporization under equilibrium conditions. The film vaporization proceeds through monolayer by monolayer irreversible removal of material from the films surface. At a particular constant temperature, the vacuum –film interface moves toward substrate with a constant velocity.

As shown in the Fig. 9, the model is in excellent agreement with the observed variations in the excess heat capacity of toluene films during vaporization. The activation energy for toluene vaporization determined from the fit of vaporization endotherms with Eq. 1 is 38±1 kJ·mol$^{-1}$ in the temperature range from 200 to 240 K (see the inset of Fig. 9), *i.e.*, in agreement with reported literature values [43] . The vaporization rate data on the Arrhenius plot in the inset were calculated using Eq. 2 as explained in Sec. III C.

*In summary*, the transformation of a stable glassy phase into ordinary supercooled liquid during rapid heating and under isothermal conditions follow zero-order kinetics with temperature dependent rate constant. Thus, a single FSC experiment results in ample and easily interpreted information on the kinetic stability of a vapor deposited non-crystalline material. Taking advantage of simple kinetics of far out of equilibrium relaxation, we tested properties of stable toluene films vapor deposited at distinct rates. The results are reported in the next sections of this article.

## C. Kinetics of transformation of stable films deposited at different rates.

Figure 10 summarizes results of FSC studies of the impact of deposition rate on stability of vapor-deposited films. The film thickness was approximatly 1.8 μm in all experiments. All samples were deposited at temperature near 112 K.

As shown in Fig. 10, lowering the deposition rate results in increase of the onset temperatures of FREs. For example, a decrease in the deposition rate from 39 to 0.6 nm·s⁻¹ leads to an increase in the $T_{FRE}$ by approximately 4 K. Although seemingly small, the change in $T_{FRE}$ even by a few Kelvins signifies substantial variations in the stability of the film due to high apparent activation energies of the transformation process. As illustrated in the inset of Fig. 10, the decrease in deposition rate from 39 to 0.6 nm·s⁻¹ leads to a decrease in the transformation rate (i.e., in the velocity of transformation front at a particular temperature) by two orders of magnitude during rapid scan. The kinetic analysis of the thermograms of stable toluene films leading to this conclusion is explained immediately below.

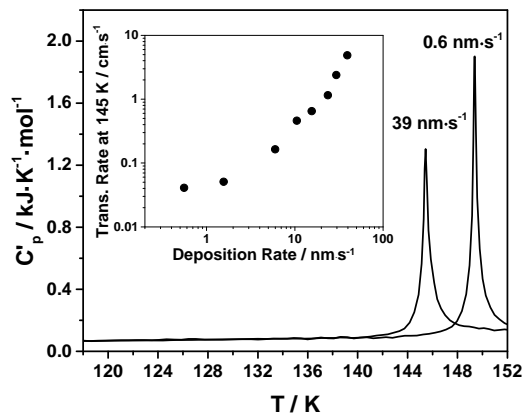

**Figure 10.** Representative thermograms of stable toluene films vapor deposited at distinct rates at 112 K. Inset shows the absolute transformation rate values at temperature of 145 K during the transformation for the films prepared at eight distinct deposition rates. The rate is reported in the transformation front velocity units (cm/s). See Fig. 11 and text for discussion.

Fig. 11, shows the absolute transformation rates of stable toluene films expressed in terms of transformation front velocities, $\vartheta(T)$, i.e., in terms of values calculated from calorimetric data under assumption that the transformation of stable toluene film into supercooled liquid starts at the film's surface and progresses into the bulk of the sample with velocity which is dependent on temperature during transformation. The $\vartheta(T)$ values are shown as function of transformation temperatures for eight films deposited at distinct rates. The deposition rate values are shown on top of the panel in Fig. 11. The absolute values of $\vartheta(T)$ were calculated from the excess heat capacity, $\Delta C_p(T)$ or $(dH/dT)$, using the following equation:

$$\vartheta(T) = \frac{\left(\frac{dH}{dT}\right) \cdot \left(\frac{dT}{dt}\right)}{\Delta H_{tr} \cdot n \cdot S} , \qquad (2)$$

where $(dT/dt)$ is the heating rate during the transformation (approximately $10^{-5}$ K·s⁻¹), $n$ is the number density of condensed toluene phase (approximately $9.4 \cdot 10^3$ mol·m⁻³), $S$ is the surface area of a film on the surface of the filament ($3.4 \cdot 10^{-7}$ m²), and $\Delta H_{tr}$ is the enthalpy change during the transformation. $\Delta H_{tr}$ values were obtained by integrating heat capacity of each film over the temperature range of transformation. The average enthalpy of transformation derived from the enthalpy curves was in the range from 2.3 to 2.7 kJ·mol⁻¹ for films deposited at rates from 39 to 0.6 nm·s⁻¹ respectively.

As shown in Fig. 11, the softening of each film of stable toluene phase follows zero order kinetics characterized by an effective activation energy of approximately 200±15 kJ·mol⁻¹. The temperature value near 145 K offers a convenient mark for comparison of transformation rates of stable toluene films because the ranges of $\vartheta(T)$ values for the films deposited at various rates overlap at this temperature. The transformation rate data near 145 K are summarized in the inset of Fig 10.

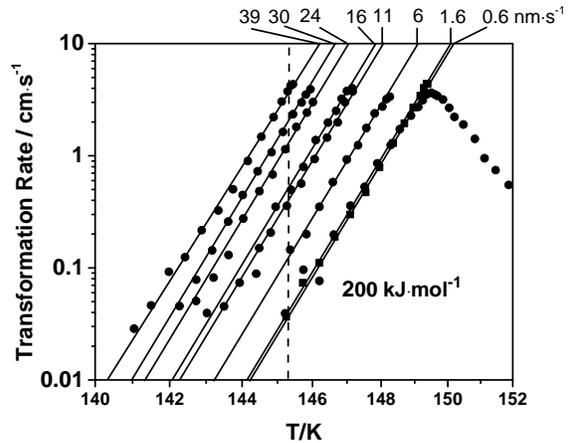

**Figure 11.** Transformation rates of stable toluene films into ordinary supercooled liquid. The films were deposited at rates varying from 0.6 to 39 nm/s at 112 K. The rate values reported in transformation front velocity units (see text for details of calculations). 145 K represents a convenient mark for comparison the isothermal transformation rates (dashed line) because the rate data ranges for the films deposited at distinct rates overlap at this temperature. The transformation rate values near 145 K are reported in the inset of Figure 11.

The strong dependence of film's kinetic stability on the deposition rate observed in our FSC experiments may seem to be in contrast to the lack thereof in AC calorimetry experiments conducted Ahrenberg *et. al.* [15]. However, we emphasize that the probe of the sample's kinetic stability employed in our FSC experiments (fast relaxation at temperatures tens of degrees above $T_g$) is drastically different from that used in AC calorimetry experiments where the transformation was monitored at temperatures no more than 5 K above the standard glass transition temperature. In short, the FSC technique may be more sensitive to relatively small variations in the structural and thermodynamic properties of stable toluene films vapor deposited at distinct rates. Note also that the deposition temperature (112 K) in our experiments is different from that in the AC calorimetry studies (105 K). It is possible that the dependence of structural and thermodynamic properties of stable toluene films on the deposition rate is stronger at higher deposition temperatures characteristic of the FSC studies (see Sec. IV B for discussion).

### D. Enthalpy of stable films deposited at different rates.

In order to gain insights into structure and formation mechanism of stable toluene films, we analyzed the relative enthalpies of the films deposited at distinct rates. Figure 12 summarizes the results of this analysis. The enthalpy values during rapid heating were obtained by integration of the thermograms over temperature in the range from 117 K to 170 K. Although the starting (deposition) temperature was near 112 K, the high noise in the current and voltage data in the beginning of the rapid scans made direct determination of enthalpy over the first few Kelvins of the thermogram's temperature axis difficult. Nevertheless, the extrapolation of enthalpy to the initial temperature was trivial due to similarities in the shapes of enthalpy curves. As shown in Fig. 12, the variations in enthalpy of stable toluene films with temperature (solid lines) are similar in the case of stable films prepared by deposition at distinct rates and can be approximated well by second order polynomials with different offsets (dotted curves).

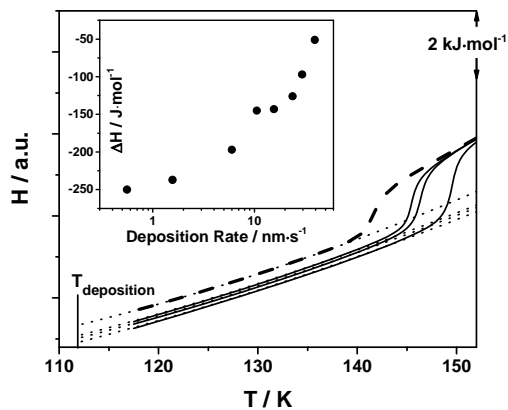

**Figure 12.** The enthalpies of selected stable toluene films vapor deposited at distinct rates during rapid heating (solid lines). The enthalpy values were inferred from the FSC thermograms by integrating heat capacity over temperatures in the range from 120 to 170 K. The enthalpy of ordinary aged (see text for details) glass is shown for comparison (dashed line). According to our estimates, the fictive temperature of ordinary glass is near 115 K. The differences between the enthalpy of ordinary glass and those of the vapor deposited samples as a function of the deposition rate are shown in the inset.

Note it is difficult to determine accurately the *absolute* enthalpy of stable toluene film. A few percent systematic uncertainties in parameters such film's mass may result in a *systematic* offset in the absolute enthalpy values on the order of a hundred of J·mol$^{-1}$ due to accumulation of errors during integration. Thus, in order to facilitate quantitative analysis of the enthalpy of vapor-deposited films, we compare it to the enthalpy of ordinary glass. The ordinary glass film was prepared by vapor deposition of toluene near 128 K, *i.e.*, when near-equilibrium relaxation times are several orders of magnitude shorter than typical deposition time and formation of stable phase is therefore impossible[31, 32]. The deposition was followed by cooling the sample to 115 K with a rate of approximately 3 K·s$^{-1}$. The film was then aged at this temperature for 10 minutes prior to FSC scan to ensure that the enthalpy of ordinary glass sample prepared for our FSC experiment is *not* higher than that of glassy sample obtained by cooling the melt in standard DSC investigations. The analysis of primary data was identical to that in the case of stable toluene samples. Note also that the thermograms of ordinary toluene films were measured on the same day as those of stable films, *i.e.*, all experimental conditions and treatment of data are identical for stable and ordinary samples. The enthalpy of the ordinary glass sample is shown in Fig. 12 as dashed curve.

The inset in Figure 12 summarizes the result of comparative analysis of enthalpy of the stable toluene films. The values along the y-axis, *ΔH*, were obtained by subtracting the enthalpy of the ordinary glassy toluene sample from those of stable films and plotted against the logarithm of the deposition rate. As shown in the inset of Fig. 12, the films deposited at the lowest rate (0.6 nm·s$^{-1}$) are stabilized by approximately 250 J·mol$^{-1}$ in respect to the glassy films prepared by standard method, *i.e.*, by cooling the melt. Note, once again, that the absolute values of enthalpy difference reported in the inset of Fig. 12 may be a subject to a systematic offset.

According to our best estimates, the enthalpy difference for the stable film deposited at the lowest rate may be as large as 300 J·mol$^{-1}$, but no less than 200 J·mol$^{-1}$ in respect to that of an ordinary sample. The lowest estimate for |ΔH| represent the case when the enthalpy of a stable film vapor-deposited at the highest rate (39 nm·s$^{-1}$) is nearly identical to the enthalpy of ordinary sample. This, however, is unlikely because the thermograms of the ordinary and the vapor deposited samples still show significant differences (approximately 2 K) in the temperature onset of fast relaxation even for the films deposited at the highest heating rates. In short, the enthalpy of the most stable vapor deposited film is either very close to the enthalpy of crystalline toluene (250 J·mol$^{-1}$ below the enthalpy of glass at 112 K) [24, 33] or below the enthalpy of the crystalline phase. As we explain in Sec III, the later suggestion may have a merit and should not be rejected outright taking into account a proposed mechanism of formation of stable vapor-deposited samples. Before we summarize and discuss in detail the findings described in the previous sections of this article, we report the results of our FSC studies of impact sample thickness and substrate morphology on stability of vapor deposited toluene films.

**E. Impact of film thickness and substrate properties.**

Past studies of highly stable glasses provided evidence that the properties of vapor deposited phase may depend on the sample size (film thickness). For example, the kinetics of transition from highly stable state to ordinary supercool liquid may change significantly as the sample thickness exceeds 1000 nanometers[4]. In order to explore the impact of the confinement on the stability of vapor deposited films of toluene, we have collected FSC thermograms for the films of thicknesses ranging from 2 μm to 25 nm. The films were vapor-deposited at 112 K with the rate of 15 nm·s$^{-1}$ at 112 K. In the case on nanometer scale films, the data from multiple FSC thermograms (up to several hundreds) were averaged to improve the signal to noise ratio. The results of these experiments are summarized in Fig. 13.

As shown in the Figure 13, the variations in the kinetics of vapor deposited toluene films are negligible at thicknesses above 250 nm. However, the stability of vapor deposited films declines dramatically when the thickness of the film is lowered from 250 to 25 nm. As shown in the inset of Fig. 13, the kinetics of transformation into ordinary liquid during rapid scan can still be approximated by Arrhenius dependence in the case of nanoscale films. The rate values were calculated using Eq. 2 as explained in Sec. III C. The apparent activation energy in the case of films of thickness of 25, 100 and 250 nm is less than a half of that in the case of films with thicknesses above 250 nm.

The sudden change in transformation kinetics at film thicknesses below 250 nm is surprising. Ahrenberg *et. al.* studied the transformation kinetics of films with the thickness of 390 nm under isothermal conditions[15]. The transformation rate of these thin films followed zero-order kinetics and remained constant until 95 % of the initially stable film was transformed into ordinary supercooled liquid, *i.e.*, until the thickness of the remaining stable phase was at most a few tens of nanometers. In other words no sudden or gradual changes in kinetics of transformation with film thickness were observed in those experiments.

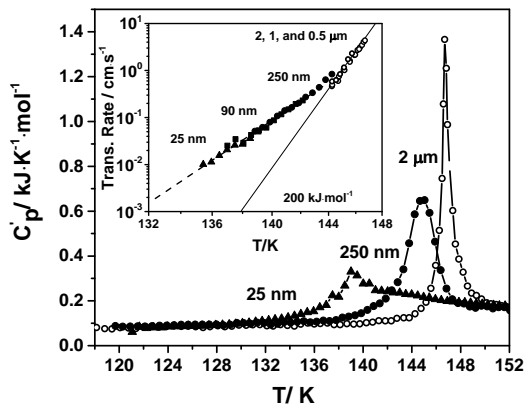

**Figure 13.** Selected FSC thermograms of vapor deposited films of toluene of distinct thicknesses. Each film was deposited near 112 K at the rate of approximately 15 nm/s. As shown in the inset, the 2, 1, and 0.5 µm films show transformation kinetics which are nearly indistinguishable and characterized by large activation barrier. The decrease in the thickness of the film to 250 nm results in sudden change in the rate and kinetics of transformation. The transformation of nanoscale films occurs with the rates orders of magnitude faster than that observed in the case of thicker films, but also may be characterized by drastically lower transformation activation energy.

In order to gain insights into fast relaxation of vapor deposited toluene films in the limit of low thicknesses, we have examined the surface of our substrate (filament) by means of Scanning Electron Microscopy (SEM). Cursory analysis of the SEM image of the filament resulted in a value on the order of 200 nm for the average surface roughness of the filament, which is likely to exceed the surface roughness of the chip calorimeter employed in the AC experiments.

The correlation between the surface roughness scale and the thickness of the film at which decline in the stability of the toluene films is observed in our experiments may be interpreted as an evidence of *kinetic anisotropy* of the stable toluene samples. Indeed, the vapor-deposits with *effective* thickness less than the surface roughness scale are likely to form preferentially at valleys on the rough surface and not cover the filament completely. In other words, such deposits are likely to be in the form of pools of stable phase with average size on the order of the surface roughness as opposed to continuous film in the case of smooth substrate. The fact that vapor deposited films on a flat substrates remain stable even when their thickness is on the order of nanometers, indicates that the lateral size of the sample is more critical for high stability than the thickness of the sample.

## IV. DISCUSSION

### A. Mechanisms of fast transformation of stable vapor-deposited phase of toluene.

According to current understanding of the mechanism of the stable vapor deposited phase transformation into ordinary supercooled liquid, the transformation rate is assumed to be controlled by molecular mobility in the liquid phase. In other words, the rate of transformation is controlled by diffusion of constituents from the solid-liquid interface into the liquid. This view is based on the theoretical analysis of Wolynes [44] supported by experimental results [15]. For example, transformation times, $t_{tr}$, measured in AC calorimetry experiments do show sub-linear correlation between $t_{tr}$, and the near equilibrium relaxation times, $\tau_{rel}$, in a temperature range from 118 K to 123 K. Furthermore, our analysis of the relaxation kinetics data of Ahrenberg et.al.[15], results in the value on the order of 180 kJ·mol$^{-1}$ for the apparent activation energy for the zero-order transformation rate constant, which is close to the expected apparent activation energy for the relaxation rates in ordinary liquid toluene in this temperature range [31, 32] (155 kJ·mol$^{-1}$).

The transformation rates measured in the FSC experiments (see Fig.11) may seem to be in contrast with past studies [15]. If the kinetics of the transformation from stable phase to ordinary supercooled liquid were governed by diffusion rates into the ordinary liquid, the activation energy for transformation near 145 K, *i.e.*, during fast transformation observed in the FSC experiments, must be on the order of the activation energy for the relaxation rate of ordinary liquid at this temperature. The apparent activation energy for relaxation rates in ordinary supercooled liquid in the temperature range from 140 to 150 K can be estimated from reported relaxation times [31, 32] and is on the order of 55 kJ·mol$^{-1}$. However, we observe a zero-order transformation kinetics which is characterized by a barrier of nearly 200 kJ·mol$^{-1}$, *i.e.*, on the order of that characteristic of the transformation observed in AC calorimetry experiments at much lower temperatures.

The unexpectedly high apparent activation energies of the transformation of stable phase observed in FSC experiments can be interpreted in two ways. First, the transformation of the stable phase into ordinary supercooled liquid may occur in two stages. During the first stage, the transformation front propagates rapidly through the stable phase leaving behind a high density liquid, which then relaxes into ordinary supercooled liquid of normal density. Thus, the first elementary step of the transformation is the diffusion into anomalously dense phase, which is essentially a low fictive temperature liquid with structure similar to that of ordinary liquid at temperatures close to its standard glass transition.

*Second*, the structure of the stable phase is unique, *i.e.*, the toluene molecules are locked into unique configuration which requires overcoming an energy barrier in order to separate from the top layer of the remaining stable phase films during transformation. An example of such a configuration could be a two-dimensional (2D) self-assembled layer or 2D crystal which may transform into ordinary liquid via highly activated process of defect formation and propagation.

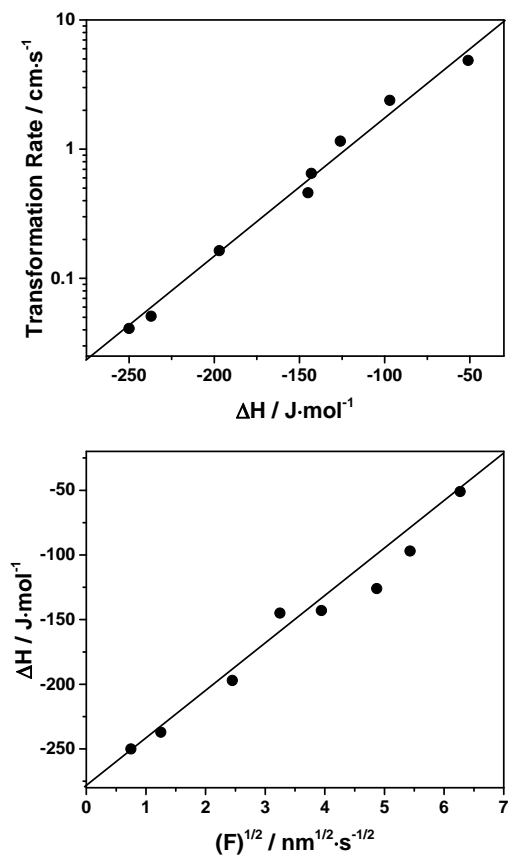

**Figure 14.** Correlations between transformation kinetics, initial relative enthalpy, and the deposition rate. *Upper panel*: The logarithm of the transformation rate of films vapor-deposited at rates ranging from 0.6 to 39 nm/s shows linear dependence on their relative initial enthalpies. *Lower panel:* The enthalpy of the films prepared at distinct rates shows a linear dependence on the square root of the deposition rate, *i.e.,* is inversely proportional to the square root of the deposition time. See text for discussion. The solid lines are eye guides, not results of linear fit.

Although the exact mechanism of fast transformation of the stable phase during rapid heating is not completely clear, it is undeniably sensitive to the initial thermodynamic and structural state of the vapor-deposited samples, and easily quantifiable (see Fig. 10 and 11). As shown in the upper panel of Figure 14, the logarithm of the transformation rate of stable toluene films vapor-deposited at distinct rates is proportional to their initial relative enthalpy. In short, the Fast Scanning Calorimetry may and should be used as a sensitive probe of structure and thermodynamics of non-crystalline phases.

**B. Mechanism of formation of stable vapor-deposited phase.**

As we mentioned in the introduction for this article, several experimental observations are not properly explained by current empirical scenario of stable phase formation which postulates that high surface diffusion rates of the phase constituents allow much faster relaxation of the sample to low enthalpy state which, in principle, can be achieved by aging the ordinary glass over extraordinary long times. For example, past wide angle X-ray scattering (WAXS)[3] studies revealed structural anisotropy in stable vapor-deposited tris-naphthylbenzene sample. As we explained in Sec. III E, the variations in stability of toluene films deposited on a rough substrate implies that the stable phase is characterized by *the kinetic anisotropy* of the sample, *i.e.*, that the lateral dimensions of the stable sample are more important for stability than the sample thickness. The formation of anisotropic (layered) stable phase was also observed in computational studies of the structure of a vapor-deposited glass of trehalose[3].

The results of the FSC experiments reported in this article, along with the WAXS, and the computational studies make it tempting to consider the mechanism of formation of stable non-crystalline phases in the context of formation and morphology of sub-monolayer aggregates of material in molecular beam epitaxy[45]. With a caveat that such an analysis is rather speculative at the present time, we offer a scenario for stable phase formation immediately below.

In Frank - van der Merwe mode, growth of vapor-deposited samples begins with nucleation of 2D islands on the surface of the substrate [45]. We assume that the deposition of toluene and other species results in formation of tightly packed and kinetically stable 2D phase which is likely to be distinct from bulk crystal or ordinary glass [21, 46, 47]. Indeed, recent WAXS and computational studies of stable toluene films indicate that, unlike ordinary glass, the vapor deposited samples may consist of a range of locally stable structures. [21]. During deposition, the islands grow until the entire surface of the substrate is covered. The film growth continues with the formation of consequent layer of 2D islands, i.e. the resulting film is a stack of 2D layers.

The island formation process outlined above, although conceptually simple, may result in wide range of island size distributions and morphologies which are sensitive to deposition conditions [45-49]. While the process is governed primarily by a simple ratio of the diffusion rate (D) to the deposition flux (F), the average isle size, the isle size distribution and the isle morphologies in a particular layer in the stack may vary dramatically depending on interplay between surface diffusion and deposition rates at a particular deposition temperature. The complex dependence of the average 2D isle size in on the deposition conditions, the stack may explain some "scattering" experimental results on kinetic stability and structure of stable vapor-deposited phases. Although fairly sophisticated theories of island formation have been developed, which take into account such as island coalescence and percolation[49], we will to try to explain the results observed in FSC calorimetry using the most basic scenario.

We assume that the *final* average island size, $L$, is determined exclusively by the surface density of island nuclei at intermediate coverage, *i.e.*, that the coalescence of the islands is impossible. In such growth regime, the density of islands, $n_{is}$, scales with the ratio of the diffusion rate, $D$, to deposition rate, $F$, as

$$n_{is} \sim \left(\frac{D}{F}\right)^{-\chi} , \qquad (3)$$

where the exponent $\chi$ is related to the critical size of the island's nucleus and can be expressed in terms of number of monomers in the nucleus, $i$, as $i/(i+2)$ [45, 48, 49]. The average size of the island, $l_{is}$, scales with the island density as $n_{is}^{-1/2}$, and *overall length of island boundaries*, $L_b$, must be proportional to $n_{is}^{1/2}$.

As a particular layer is formed, the average size of the island defines the enthalpy of the monolayer and, ultimately, the enthalpy of the entire film. The boundaries between islands may be viewed as "defect sites" [50, 51] which contribute to increase in the enthalpy of the entire sample. The number of the defects must be proportional to the net span of island boundaries. Thus, the enthalpy variations of a stable film must scale as

$$\Delta H \sim n_{is}^{1/2} \sim \left(\frac{F}{D}\right)^{\frac{i}{2(i+2)}} , \qquad (4)$$

Note that Eq. 4 simplifies if more than 10 monomers are required to form a stable critical nucleus of an island at particular deposition temperature. In other words, the enthalpy of the stable films is likely to be a linear function of $t_{dep}^{-1/2}$, where $t_{dep}$ is the deposition time, or a linear function of $F^{1/2}$. The lower panel of Fig. 14 shows the relative enthalpies of the stable toluene films vapor-deposited at various rates plotted against the square root of the deposition rate. Within the errors of determination, the enthalpy of vapor deposited samples conforms well to the linear function in these coordinates. This observation agrees with and explains the results of AC calorimetry study of vapor deposited toluene films, where the fractional heat capacity decrease at low deposition rate seemed to plateau at low deposition rates[15]. Indeed, any sub-linear dependence may appear to plateau in the at large values of the argument.

As we already mentioned, the conjecture proposed in this section is not immediately proven by the simple scaling relationship between enthalpy and the deposition rate which we observe in our FSC experiments. However, it provides basis for reconciling some of the contrasting results obtained by different researchers. For example, it provides possible explanation for decrease in kinetic stability of toluene films in the limit of low deposition rates observed by Leon- Gutierrez *et. al.* [5]. Indeed, thin films are inherently unstable [52-54]. Any structural mismatch between the topmost islands and underlying films is likely to grow with island size. Thus, at low deposition rates, the islands may reach a critical size when collapse into 3D clusters is thermodynamically favored[54]. Although the mechanism and structure of stable vapor-deposited phases may yet be proven different from those proposed here, we are certain that any discussion of the surface facilitated formation and softening of stable vapor-deposited non-crystalline phase would benefit from taking into account the wealth of information from the field of classical molecular beam epitaxy despite the non-crystalline nature of the materials formed.

## V. SUMMARY AND AFTERWORD

Fast scanning calorimetry is extremely sensitive to structure of vapor –deposited non-crystalline phases and can be used to characterize simultaneously their kinetic stability and thermodynamic stability. Transformation of stable vapor deposited samples of toluene during heating with rates in excess $10^5$ K·s$^{-1}$ follows the zero-order kinetics with the effective activation energy of the order of 200 kJ·mol$^{-1}$, i.e., is highly activated and independent of the diffusion rates in ordinary supercooled liquid at the same temperatures. The transformation rate, however, shows strong correlation with the initial enthalpy of the sample, which decreases with deposition time according to sub-linear law. Studies of the kinetics of stable toluene films with nanoscale thicknesses reveal a sudden increase in the transformation rate and change in transformation kinetic for films thinner than 250 nm which correlates with the substrate roughness scale. The correlation between the substrate roughness and its stability implies kinetic anisotropy, i.e., implies that the lateral dimensions of the stable sample have a greater impact on its stability than the film's thickness. These findings indicate that the stable vapor deposited films of non-crystalline toluene may represent a stacked phase which properties are determined by 2D island growth kinetics during in Frank - van der Merwe deposition mode. Further experiments are needed, of cause, to verify this conjecture.

## ACKNOWLEDGMENTS

This work was supported by the US National Science Award 1012692. We are grateful to Liam O'Reilly for assistance with some measurements. We are also grateful to Ulyana S. Cubeta for her tests and analysis of the temperature calibration procedure used in the FSC experiments.